\documentclass[useAMS,usenatbib,referee]{biom}
\pdfoutput=1

\usepackage{graphicx,multirow,rotating, inputenc}

\def\bSig\mathbf{\Sigma}

\title[Joint models FPLS]{\bf 
Partial Least Squares for Functional Joint Models\\
\large with applications to the Alzheimer’s Disease Neuroimaging Initiative Study}

\author{Yue Wang \\
Department of Biostatistics, University of Washington, Seattle, Washington, U.S.A.
\and Joseph G. Ibrahim, 
Hongtu Zhu \\
Department of Biostatistics, University of North Carolina at Chapel Hill, Chapel Hill, North Carolina, U.S.A.}

\begin{document}





\pagerange{\pageref{firstpage}--\pageref{lastpage}} 
\volume{xx}
\pubyear{2019}
\artmonth{November}


\doi{xxxxxxxx.x}


\label{firstpage}


\begin{abstract}
Many biomedical studies have identified important imaging biomarkers that are associated with both repeated clinical measures and a survival outcome. The functional joint model (FJM) framework, proposed in \cite{LiLuo2017}, investigates the association between repeated clinical measures and survival data, 
 while adjusting for both high-dimensional images and low-dimensional covariates based upon the functional principal component analysis (FPCA). 
 In this paper, we propose a novel algorithm for the estimation of FJM based on the functional partial least squares (FPLS). Our numerical studies demonstrate that, compared to FPCA, the proposed FPLS algorithm can yield more accurate and robust estimation and prediction performance in many important scenarios.
We apply the proposed FPLS algorithm to the Alzheimer's Disease Neuroimaging Initiative (ADNI) study. Data used in the preparation of this article were obtained from the ADNI database.
\end{abstract}

%



\maketitle


%

\section{Introduction}
\label{s:intro}
Many prospective cohort studies and clinical trials investigating neurodegenerative diseases such Alzheimer's disease (AD) collect repeated measurements of clinical variables, event history and biomedical imaging data.
A motivating example is the Alzheimer's Disease Neuroimaging Initiative (ADNI) study. ADNI currently has  4 phases: ADNI1, ADNI-GO, ADNI2 and ADNI3, and the primary goal is to test whether serial magnetic resonance imaging (MRI), positron emission tomography (PET) and neuropsychological assessments can be used to measure the progression of AD. Participants were assessed at multiple visits. At each visit, various clinical measures, brain images and neuropsychological assessments were collected. Detailed information about ADNI  can be found on the official website ``http://www.adni-info.org".

Since mild cognition impairment (MCI) is known as a transitional stage between normal cognition and AD, it is of great interest to predict the progression from MCI to AD.
Thus, we selected 236 patients who had been diagnosed with MCI from ADNI1 without missing data in the covariates of interest. The selected patients had at least one Alzheimer's Disease Assessment Scale-Cognitive (ADAS-Cog) score after the baseline measurement. The ADAS-Cog score measures cognition functions and ranges from 0 to 70, with a higher score indicating poorer cognitive function. We consider the AD diagnosis as the survival event of interest. Among the 236 individuals, 92 individuals progressed to AD before the completion of ADNI1 and the remaining 144 individuals did not. Thus, the time of conversion from MCI to AD can be treated as right censored time-to-event data and the censoring is non-informative about the progression to AD. Demographic information of the selected 236 patients and summary statistics of the ADAS-Cog score can be found in the supporting information. 

\begin{figure}
\centerline{\includegraphics[width=\textwidth,height=0.6\textheight]{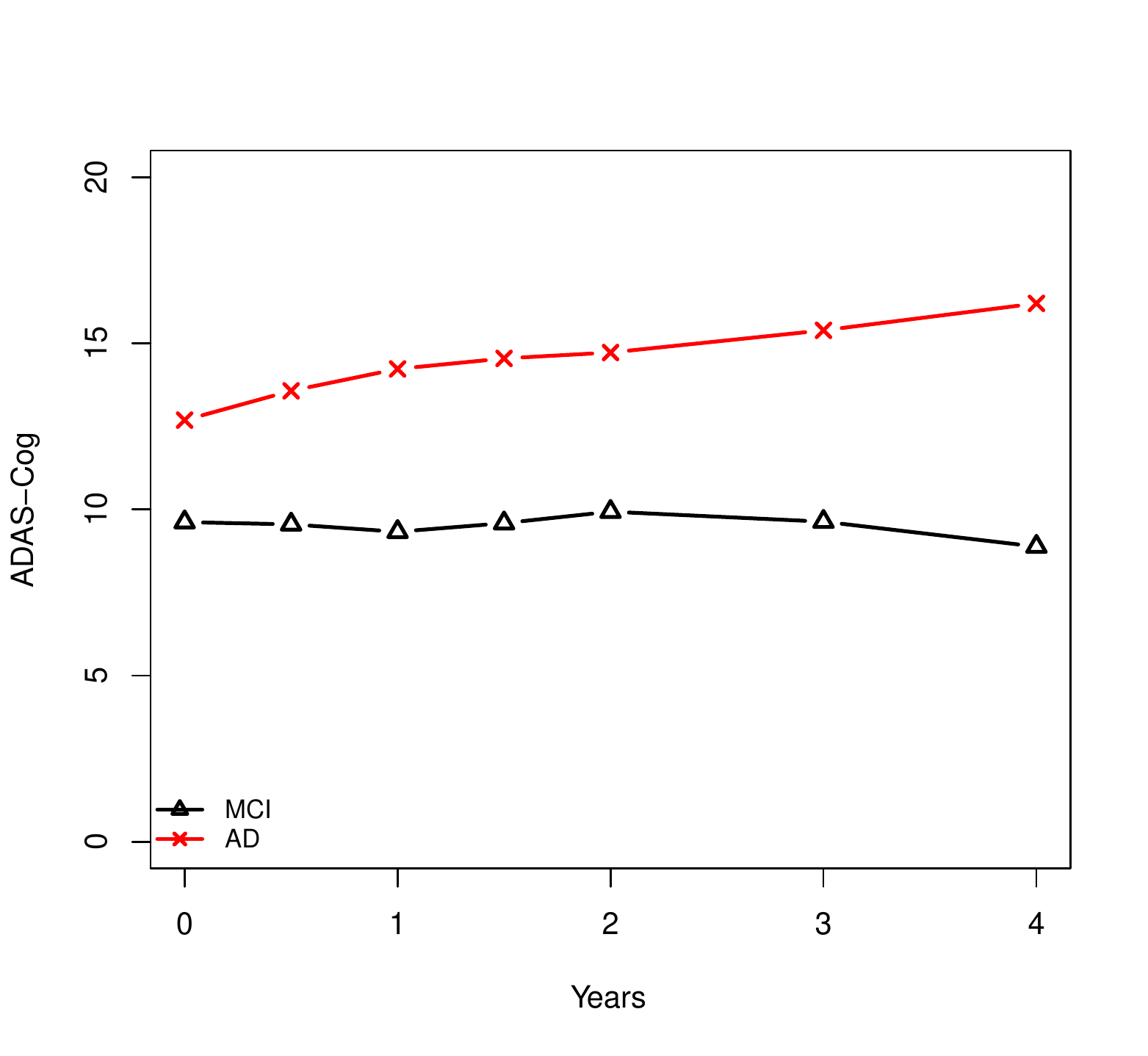}}
\caption{The ADAS-Cog score at year 0, 0.5, 1, 1.5, 2, 3 and 4 for both MCI and AD patients. At each time point, the score is averaged on those subjects who have not dropped out before this time point.}
\label{intro_fig1}
\end{figure}

Figure \ref{intro_fig1} displays the average ADAS-Cog score at each follow-up time separately for the MCI and AD group. It can be seen that the ADAS-Cog score increases with time for the AD group, whereas for the MCI group, the trend is not evident. The AD group tends to have higher ADAS-Cog scores, indicating that the ADAS-Cog score may be predictive to the progression of AD. Moreover, a lot of existing work reported the association between brain imaging predictors with the progression of AD. For example, AD and MCI patients were shown to have $27\%$ and $11\%$ smaller hippocampal volumes respectively as compared with normal controls \citep{Du2001}. \cite{LeeZhu2015} and \cite{kong2018} both demonstrated the predictive value of the hippocampus surface data to the progression of AD. 

Thus, it is of great interest 
to examine the association between longitudinal ADAS-Cog scores and the progression of AD while adjusting for high-dimensional imaging predictors. 
A recent work by \cite{LiLuo2017} proposed a functional joint model (FJM) for this purpose. FJM integrates functional regression models with joint models of longitudinal and time-to-event data. There is a rich literature in both areas. For instance, in the past decades, generalized functional linear models (GFLM) have been widely discussed and applied in various areas including finance, biology and sociology. Please see \cite{Cardot1999,Cardot2003}, \cite{Muller2005} and the references therein for an extensive review of GFLM. Then, \cite{YaoMuller2005} extended GFLM to longitudinal data, and \cite{LeeZhu2015} and \cite{kong2018} both extended GFLM to the proportional hazards model. Some of the earliest work on joint models of longitudinal and time to event data is in \cite{TsiatisDeGruttola1995} and \cite{wulfsohn1997}.


The major challenge of fitting a functional model, including FJM, is the ultrahigh dimensionality of the functional predictor. 
A common strategy is to find the ``best" low-dimensional features that approximate the high-dimensional functional predictor, which is also known as  dimension reduction. The functional principal component analysis (FPCA) has been a popular dimension-reduction tool for functional data over decades. 
See an extensive review and application of FPCA  in \cite{Besse1986}, \cite{RamsayDalzell1991}, \cite{Boente2000335}, \cite{James2000}, \cite{Hall2006}, \cite{muller2010}, \cite{LiLuo2017} and the references therein. Based upon the eigen-decomposition of the covariance kernel of the functional data, FPCA finds the low-dimensional space that preserves most of the variation of the functional data; that is, the space spanned by top eigenfunctions. While FPCA serves as a promising tool for exploring functional data, there are two major concerns when it comes to regression. First, as the unknown slope function may not necessarily lie in the space spanned by top eigenfunctions, the estimation and prediction accuracy may suffer if the number of eigenfunctions used for regression is underestimated. Second, it is well known that consistent estimation of  eigenvectors is highly challenging in ultrahigh-dimensional settings \citep{Jung2009}, especially for tail eigenvectors. These two concerns bring up a natural question: if (part of) the unknown slope function lies in the space spanned by some tail eigenfunctions, which cannot be accurately estimated due to the limited sample size, is there an alternative and better approach to estimate the unknown slope function?

To address this question, we consider the functional partial least squares (FPLS), first proposed by \cite{Preda2007} for functional linear models (FLM). Compared to FPCA, FPLS has two major advantages. First, FPLS does not depend on accurate estimates of eigenfunctions. Second, FPLS incorporates information from the outcome so that the top FPLS basis functions are always the most predictive to the outcome. 
However, due to its computational and theoretical complexity, FPLS has not gained enough popularity until the recent work by \cite{Hall2012} that proposes a simplified FPLS algorithm for FLM, called alternative PLS (APLS). Through numerical studies,  \cite{Hall2012} demonstrates that APLS can capture the interaction between the functional predictor and the outcome using a fewer number of components than FPCA.

Motivated by the encouraging performance of APLS, in this article, we propose an FPLS algorithm for the estimation and prediction of
FJM. Our specific contributions include: (i) We extend APLS to the complex FJM framework in a rigorously mathematical manner; (ii) we show by numerical studies that the proposed FPLS algorithm can yield considerably more accurate, robust estimation and prediction results than FPCA in many important scenarios and (iii) unlike the FPCA-based algorithm in \cite{LiLuo2017} that only handles baseline imaging data, our FPLS algorithm can deal with longitudinal imaging data; this allows to dynamically predict disease progression. 
The rest of the paper is organized as follows.  
Section \ref{JMest} introduces the FPLS algorithm for FJM. Section 3 discusses the details of the implementation of the FPLS algorithm.
Section 4 presents a simulation study with ultrahigh-dimensional images in various settings.
 Section 5 presents a thorough analysis of the selected 236 ADNI patients.
 Concluding remarks and discussions are presented in Section 6. 

\section{Methods}
\label{JMest}

In this section, we first introduce the FJM proposed by \cite{LiLuo2017}.
Then, we review APLS for FLM proposed by \cite{Hall2012}. Next, we extend APLS for FJM. 

\subsection{FJM}
For each subject $i = 1, \ldots, n$ at visit $k = 1, \ldots, K_i$, 
we observe $\{y_{ik}, {\bf  x}_i, {\bf  z}_{ik}, \bmath \omega_i, T_i, \Delta_i \}$, where $y_{ik}$ is the outcome of interest observed at time $t_{ik}$, ${\bf z}_{ik}$ is a $p_z \times 1$ vector observed at $t_{ik}$ and $\bmath \omega_i$ is a $p_\omega \times 1$ vector of time-invariant covariates, which may overlap with ${\bf z}_{ik}$. The ${\bf x}_{i}=(x_{i}(s):  s\in \mathcal S)$  is  baseline imaging ( functional) data observed 
at a set of grid points in an nondegenerate,  compact space $\mathcal S\subset R^K$ for the $i$-th subject, where $K>0$ is a positive integer. The $T_i$ is the observed survival time, defined as $T_i=\min\{T_i^*,C_i\}$, where $T_i^*$ and $C_i$ are,  respectively,  the true event time and censoring time. The event is observed if $\Delta_i = 1$, and censored otherwise. It is assumed throughout the paper that the censoring mechanism is independent of the true event process.

The FJM in \cite{LiLuo2017} consists of a longitudinal model and a survival model. The longitudinal model is given by 
\begin{equation}\label{intro:model:jm:long}
 y_{ik} =  m_i(t_{ik}) + \epsilon_{ik} = \beta_0 + {\bf z}_{ik}^{T}\bmath  \beta_1+ \bmath  q_{ik}^T \bmath  u_i + \int_{\mathcal{S}}x_i(s)b_0(s)ds + \epsilon_{ik},  
\end{equation}
where $m_i(t_{ik})$ is the unobserved true longitudinal trajectory of the $i$-{th} subject at $t_{ik}$, $\bmath q_{ik}$ is a subset of ${\bf z}_{ik}$, $\beta_0$ is the intercept, $\bmath \beta_1$ is a $p_z \times 1$ vectors of regression coefficients, and $b_0(s)$ is the functional parameter that characterizes the association between $x_i(s)$ and $y_{ik}$. We also assume $\bmath u_i \stackrel{i.i.d}{\sim} N(0, \Sigma_u)$ and $\epsilon_{ik} \stackrel{i.i.d}{\sim} N(0, \sigma_{\epsilon}^2)$.  The survival model is given by
\begin{equation}\label{intro:model:jm:surv}
\lambda_i(t|\bmath  \omega_i, x_i(s), \bmath u_i)=  \lambda_0(t)\exp(\bmath  \omega_i^T\bmath{\gamma} + \int_{\mathcal{S}}x_i(s)b_1(s)ds + \alpha m_i(t)), 
\end{equation}
where $\lambda_0(t)$ is an unknown baseline hazard function, $\bmath \gamma$ is a $p_\omega \times 1$ vector of regression coefficients,  and $b_1(s)$ is the functional parameter that characterizes the association between $x_i(s)$ and the survival outcome. The scalar parameter $\alpha$ quantifies the association between the true longitudinal trajectory and the survival process at the same time.  

\begin{remark}
We discuss our assumptions on  FJM 
(\ref{intro:model:jm:long}) and (\ref{intro:model:jm:surv}). First, the longitudinal marker $y_{ik}$ in (\ref{intro:model:jm:long}) is assumed to be normally distributed, which is a standard assumption in the literature. The association between the longitudinal and time-to-event processes is represented by the proportional hazards model (\ref{intro:model:jm:surv}), in which $m_i(t)$ is assumed to be continuous. For definiteness, the conditional hazards function $\lambda_i(t|\bmath  \omega_i, x_i(s), \bmath u_i)$ is taken to depend linearly on the longitudinal marker through the current $m_i(t)$. 
\end{remark}
\begin{remark}
The FJM (\ref{intro:model:jm:long}) and (\ref{intro:model:jm:surv}) can be extended to a more general FJM framework as follows:
\begin{equation}\label{intro:model:jm:long:gen}
 y_{ik} =  m_i(t_{ik}) + \epsilon_{ik} = \beta_0 + {\bf z}_{ik}^{T}\bmath  \beta_1+ \bmath  q_{ik}^T \bmath  u_i + \int_{\mathcal{S}}x^{(l)}_i(s,t_{ik})b_0(s)ds + \epsilon_{ik}
\end{equation}
and 
\begin{equation}\label{intro:model:jm:surv:gen}
\lambda_i(t|\bmath  \omega_i, x_i(s), \bmath u_i)=  \lambda_0(t)\exp\left(\bmath  \omega_i^T\bmath{\gamma} + \int_{\mathcal{S}}x^{(s)}_i(s)b_1(s)ds + \alpha m_i(t)\right).
\end{equation}
Note that compared with (\ref{intro:model:jm:long}), 
the longitudinal model (\ref{intro:model:jm:long:gen}) involves longitudinal imaging data. This extension is appealing in many applications because it allows the dynamic prediction of disease progression. It is worth noting that the FPLS algorithm that will be introduced later can be seamlessly applied to this general FJM framework (\ref{intro:model:jm:long:gen}) and  (\ref{intro:model:jm:surv:gen}). But to simplify the notation, we stick to the FJM (\ref{intro:model:jm:long}) and  (\ref{intro:model:jm:surv}) in the following sections for illustrating our key ideas.
\end{remark}

\subsection{The APLS algorithm}

We review APLS proposed by \citet*{Hall2012} for the following FLM:
\begin{equation}\label{RAPLS1}
y = a_0 + \int_{\mathcal{S}}x(s)b(s)ds + \epsilon, 
\end{equation}
where $\epsilon$ is a scalar random variable with $E(\epsilon| {\bf x})=0$, $a_0$ is an  intercept, and $b(s)$ is an unknown coefficient function.
 Let  $K(\psi)(t) = \int_{\mathcal{S}}K(s,t)\psi(s)ds$ be a functional operator where $K(s,t) = \rmn{Cov}\left(x(s), x(t)\right)$. The first $p$ APLS basis functions can be constructed as $K(b), \ldots,  K^p(b)$, where $K^{j+1}(b)(s) = \int_{\mathcal{S}} K^j(b)(t)K(s,t)dt$. Compared to the FPCA basis functions, i.e., the eigenfunctions of $K(s,t)$, the APLS basis functions have two important features. First, the APLS basis functions involve the unknown parameter $b(s)$, indicating that the APLS basis functions incorporate information from both the functional predictor and the outcome. Second, the APLS basis functions are not orthonormal. To see the latter feature clearly, consider an example that $x(s) = \sum_{j=1}^\infty \sqrt{\theta_j} \xi_{j} \phi_j(s)$, where $\xi_j \stackrel{i.i.d}{\sim} N(0,1)$. The $\theta_1 \geq \theta_2 \cdots \geq 0$
is a sequence of non-increasing eigenvalues and $\phi_1(\cdot), \phi_2(\cdot), \ldots$ are their corresponding orthonormal eigenfunctions. It can be seen that $K(s,t) = \rmn{Cov}(x(s), x(t)) = \sum_{j=1}^\infty \theta_j \phi_j(s) \phi_j(t)$. Write $b(s) = \sum_{j=1}^\infty b_j \phi_j(s)$, where $b_j = \int_{\mathcal{S}}b(s)\phi_j(s)ds$. Then, it can be seen that the $p$-{th} APLS basis function is given by $K^p(b)(s) = \sum_{j=1}^\infty \theta_j^p b_j \phi_j(s)$. Therefore, it can be checked that $\int_{\mathcal{S}} K^k(b)(s) K^l(b)(s)ds = \sum_{j=1}^\infty \theta_j^{k+l} b_j^2 > 0$ for any $k \neq l$, indicating that any two APLS basis functions are not orthogonal to each other. 

Although $b(s)$ is  unknown, a consistent estimator of  $K(b)= \int_{\mathcal{S}}K(s,t)b(s)ds$ is given by 
\begin{equation}\label{APLS6}
\widehat{K(b)}(s) =  {n}^{-1}\sum_{i=1}^n\{x_i(s) - \overline{x}(s)\}\{y_i - \overline{y}\}, 
\end{equation}
where $\overline{y}=n^{-1}\sum_{i=1}^n y_i$ and $\overline{x}(s) = n^{-1}\sum_{i=1}^n x_i(s)$.  Then, we can sequentially estimate all APLS basis functions by using 
$\widehat{K^{j+1}(b)}(t) = \int_{\mathcal{S}}\widehat{K^{j}(b)}(s)\widehat{K}(s,t)ds$ for $j\geq 1$ where $\widehat{K}(s,t) = n^{-1}\sum_{i=1}^n\{x_i(s) - \overline{x}(s)\}\{x_i(t) - \overline{x}(t)\}.$ Then, an estimator of $b(s)$ in (\ref{RAPLS1}) using $p$ APLS basis functions is given by $\widehat{b}_p(s) = \sum_{j=1}^p \widehat{t}_j \widehat{K^{j}(b)}(s),$ where
\begin{equation}\label{apls:emp}
(\widehat{t}_1, \ldots, \widehat{t}_p) = \rmn{argmin}_{t_1, \ldots, t_p} \sum_{i=1}^n\big\{y_i - \overline{y} - \sum_{j=1}^p t_j \int_{\mathcal{S}}\widehat{K^j(b)}(s)[x_i(s) - \overline{x}(s)]ds\big\}^2.
\end{equation}

\subsection{The RAPLS algorithm}
The APLS algorithm cannot be directly applied to FJM (\ref{intro:model:jm:long}) and (\ref{intro:model:jm:surv}) due to two reasons. First, it does not  adjust for additional scalar covariates. Second, the estimator $\widehat{K}(b)(s)$ in (\ref{APLS6}) is not a good estimator of $K(b)(s)$ when the the relationship between $x(s)$ and $y$ is nonlinear. To bridge the gap, we first extend the APLS algorithm to the following model: 
\begin{equation}\label{RAPLS7}
y_i = {\bf z}_i^T {\bmath  \alpha} + \int_{\mathcal{S}}x_i(s)b(s)ds + \epsilon_i, 
\end{equation}
where $ {\bmath  \alpha}$ is a $p_z\times 1$ vector  of regression coefficients and  $\epsilon_i$ is an error term with  $E(\epsilon_i|{\bf x}_i,  {\bf z}_i) = 0$ and $E(\epsilon_i^2|{\bf x}_i,  {\bf z}_i) < \infty$. It is assumed that 
$ {\bf z}_i$ includes a constant $1$.   To estimate $b(s)$ and ${\bmath  \alpha}$, we develop a residual-based APLS (RAPLS) algorithm. Specifically, model (\ref{RAPLS7}) can be rewritten in the following matrix form:
  \begin{equation} \label{RAPLS8}
{\bmath  Y}   ={\bmath  Z} {\bmath  \alpha}+ \int_{\mathcal{S}}{\bmath  X}(s)b(s)ds+\bmath  \epsilon, 
\end{equation} 
where $\bmath  \epsilon=(\epsilon_1, \ldots, \epsilon_n)^T$, ${\bmath  Y}=(y_1,\ldots,y_n)^T$,  ${\bmath  Z}=[{\bf z}_1, \ldots, {\bf z}_n]^T$, and ${\bmath  X}(s) = (x_1(s), \ldots, x_n(s))^T$. 
Let $H_Z = {\bmath  Z}({\bmath  Z}^T{\bmath  Z})^{-1}{\bmath  Z}^T$ be  the orthogonal projection matrix  onto the column space of $\bmath Z$
and $M_Z= I_n - H_Z$, where $I_n$ denotes an $n\times n$ identity matrix. 
By multiplying both sizes of (\ref{RAPLS8}) by $M_Z$, we have 
\begin{equation} \label{RAPLS9}
{\bmath  Y}_Z^{\bot}=M_Z {\bmath  Y}   = \int_{\mathcal{S}}M_Z{\bmath  X}(s)b(s)ds+M_Z\bmath  \epsilon= \int_{\mathcal{S}}{\bmath  X}_Z^{\bot}(s)b(s)ds+\bmath  \epsilon_Z^{\bot},    
\end{equation} 
where $\bmath  \epsilon^{\bot}_Z=M_Z\bmath  \epsilon$. Model (\ref{RAPLS9}) can be regarded as  a special case of model (\ref{RAPLS1}) with the response vector ${\bmath  Y}_Z^{\bot}$ and the functional covariate ${\bmath  X}_Z^{\bot}(s)$.   
In this case, we need to introduce a new functional operator  as follows: 
  \[K_Z^{\bot}(\psi)(t) =\int_{\mathcal{S}}K_Z^{\bot}(s,t)\psi(s)ds= \int_{\mathcal{S}}[K(s,t)-\mbox{tr}\{[E({\bf z}^{\otimes 2})]^{-1}E[{\bf z}x(t)]E[{\bf z}^Tx(s)]\}]\psi(s)ds. \]

The operator $  K_Z^{\bot}(\psi)(t)$ corrects  the correlation between ${\bf z}$ and $x(s)$. If ${\bf z}$ and $x(s)$ are independent, then 
$  K_Z^{\bot}(\psi)$ reduces to $K(\psi)$.  Therefore, the first $p$  RAPLS basis functions for  model (\ref{RAPLS9}) are given by 
$ (K_Z^{\bot})(b), \ldots,  (K_Z^{\bot})^p(b). 
$ 
  A consistent estimator of $(K_Z^{\bot})(b)(t)$ is given by 
$\widehat{(K_Z^{\bot})(b)}(t) = n^{-1} {\bmath  Y}_Z^{\bot T} {\bmath  X}_Z^{\bot}(t).$
 
Then, we can sequentially estimate all RAPLS basis functions by using 
\[\widehat{(K_Z^{\bot})^{j+1}(b)}(t) = \int_{\mathcal{S}}\widehat{(K_Z^{\bot})^{j}(b)}(s)\widehat{K_Z^{\bot}}(s,t)ds
~~~\mbox{for}~~~ 1 \leq j\leq p-1. \]
where $\widehat{K}^\bot_{Z}(s,t) = n^{-1} \bmath X_Z^{\bot T}(s)\bmath X_Z^{\bot T}(t)$. Given $\widehat{(K_Z^{\bot})(b)}(t), \ldots, \widehat{(K_Z^{\bot})^{p}(b)}(t)$, we define 
\[(\widehat{t}_{1,\bf z}, \ldots, \widehat{t}_{p,\bf z}) = \rmn{argmin}_{(t_1, \ldots, t_p)}||{\bmath  Y}_Z^\bot - \sum_{j=1}^p t_j \int_{\mathcal{S}} \widehat{(K_Z^{\bot})^{j}(b)}(t) {\bmath  X}_Z^{\bot}(t) ds||^2,\]
where $||{\bf a}||^2 = {\bf a}^T{\bf a}$ for any column vector ${\bf a}$. Therefore, one can estimate $b(s)$ and $\bmath \alpha$ respectively by
 $\widehat{b}_{p}(s) = \sum_{j=1}^p \widehat{t}_{j,\bf z}\widehat{(K_Z^{\bot})^{j}(b)}(s)$  and $\widehat{\bmath  \alpha}_p = ({\bmath  Z}^T{\bmath  Z})^{-1}{\bmath  Z}^T[{\bmath  Y} - \int_{\mathcal{S}}{\bmath  X}(s)\widehat{b}_{p}(s)ds].$


\subsection{The FPLS algorithm for FJM}

We develop an FPLS algorithm for the FJM (\ref{intro:model:jm:long}) and (\ref{intro:model:jm:surv}) in an iterative way by integrating RAPLS with the iterative reweighted least squares (IRLS, \citealp{green1984}).  Before discussing the details, we introduce some additional notation. Let ${\bmath y_i} = (y_{i1}, \ldots, y_{iK_i})^T$, ${\bf   z}_i = ({\bf   z}_{i1}, \ldots,{\bf   z}_{iK_i})$ and $\bmath q_i = (\bmath   q_{i1}, \ldots,\bmath   q_{iK_i})$. We define $\bmath \delta_i = \bmath q_i^T \bmath u_i + \bmath \epsilon_i$ where $\bmath \epsilon_i = (\epsilon_{i1}, \ldots, \epsilon_{iK_i})^T$. Let $\bmath   1_{K_i}$ denote a $K_i\times 1$ vector of ones 
and $\Lambda_0(t) = \int_0^t \lambda_0(u)du$ denote the cumulative baseline hazard function. 
Denote $\bmath \Theta = \{\beta_0, \bmath \beta_1, \bmath \gamma, \alpha, \bmath \Sigma_u, \sigma_\epsilon^2\}$ and let $p_0$ and  $p_1$ be the number of RAPLS basis functions used for the estimation of $b_0(s)$ and $b_1(s)$, respectively. We further denote 
$\bmath\Theta_{p_0, p_1}^{(m)}, b_{0,p_0}^{(m)}(s), b_{1,p_1}^{(m)}(s)$ and $\Lambda_{0, p_0, p_1}^{(m)}(\cdot)$ as the evaluation of $\bmath\Theta, b_0(s), b_1(s)$ and $\Lambda_0(\cdot)$ respectively at the $m$-{th} iteration for $m \geq 1$.
In the following discussion, we elaborate on how to obtain $\bmath\Theta_{p_0, p_1}^{(m)}, b_{0,p_0}^{(m)}(s), b_{1,p_1}^{(m)}(s)$ and $\Lambda_{0, p_0, p_1}^{(m)}(\cdot)$ given their values at previous iteration for  $m \geq 1$ in three steps.

\textbf{Step 1:}
We first rewrite model (\ref{intro:model:jm:long}) as follows: 
\begin{equation}\label{longwls:vec}
\bmath   y_i = \bmath   1_{K_i}\beta_0 + {\bf   z}_i^{T}\bmath   \beta_1 + \bmath   1_{K_i} \int_{\mathcal{S}} x_i(s)b_0(s)ds + \bmath   \delta_i.
\end{equation}
It can be seen that $\rmn{Cov}(\bmath   \delta_i|{\bf   z}_i, x_i(s), \bmath q_i) = \bmath   q_i\Sigma_u\bmath   q_i^{T} + \sigma_\epsilon^2 I_{K_i}$, referred to as $\bmath V_i$ hereafter. Hence, given $\bmath \Theta^{(m)}_{p_0, p_1}$, $\bmath V_i$ can be evaluated as $\bmath V_i^{(m)}$ at the $m$-{th} iteration. 
By multiplying $   \left\{\bmath V_i^{(m)}\right\}^{-1/2}$ to the left on both sides of (\ref{longwls:vec}), (\ref{longwls:vec}) can be reformulated as an ordinary least squares (OLS) problem, given by
\begin{equation}\label{longols}
 \left\{\bmath V_i^{(m)}\right\}^{-1/2}\bmath y_i =   \left\{\bmath V_i^{(m)}\right\}^{-1/2}(\bmath 1_{K_i}\beta_0 + {\bf   z}_i^T\bmath \beta_1 )+ 
   \left\{\bmath V_i^{(m)}\right\}^{-1/2}\bmath 1_{K_i} \int_{\mathcal{S}} x_i(s) b_0(s)ds 
+  \left\{\bmath V_i^{(m)}\right\}^{-1/2}\bmath \delta_i.
\end{equation} 
Note that (\ref{longols}) has the same form as (\ref{RAPLS7}), and thus we can calculate the top $p_0$ RAPLS basis functions for (\ref{longols}). To simplify the notation, we denote these RAPLS basis functions by $\psi^{(m)}_1(s), \ldots, \psi^{(m)}_p(s)$. Then, model (\ref{intro:model:jm:long}) can be approximated by
\begin{equation}\label{jm:long:step1}
 y_{ik} = \beta_0 + {\bf z}_{ik}^{T}\bmath  \beta_1+ \bmath  q_{ik}^T \bmath  u_i + \sum_{j=1}^{p_0} b_{0,j} \left(\int_{\mathcal{S}} x_i(s) \psi^{(m)}_j(s)ds \right) + \epsilon_{ik}, 
\end{equation}
and $b_0(s)$ can be approximated by $\sum_{j=1}^{p_0} b_{0,j} \psi_j^{(m)}(s)$.

\textbf{Step 2:} 
We define $\mu_i = \Lambda_0(T_i)\rmn{exp}\left( \bmath   \omega_i^T \bmath   \gamma + \int_{\mathcal{S}} x_i(s)b_1(s)ds + \alpha m_i(T_i))\right)$ for $i = 1, \ldots, n$.  Given $\bmath\Theta_{p_0, p_1}^{(m)}, b_{1,p_1}^{(m)}(s)$ and $\Lambda_{0, p_0, p_1}^{(m)}(\cdot)$, $\mu_i$ still cannot be evaluated because of the unobserved random effects $\bmath u_i$ in $m_i(T_i)$. To deal with this, for any integrable function $h(\bmath u_i)$, we define 
\[E_i^{(m)}\{h(\bmath u_i)\} = E\big\{h(\bmath u_i)|\bmath y_i, T_i, \Delta_i, \bmath\Theta_{p_0, p_1}^{(m)}, b_{0,p_0}^{(m)}(s), b_{1,p_1}^{(m)}(s),\Lambda_{0, p_0, p_1}^{(m)}(\cdot)\big\}\]
as the the posterior expectation of $h(\bmath u_i)$ at the $m$-{th} iteration. Then, we can evaluate $\mu_i$ as:
\[\mu_i^{(m)} = \Lambda_{0,p_0,p_1}^{(m)}(T_i) E_i^{(m)}\left\{\rmn{exp}\left( \bmath   \omega_i^T \bmath   \gamma + \int_{\mathcal{S}} x_i(s)b_1(s)ds + \alpha m_i(T_i))\right)\right\}.\]
Following IRLS, for $i = 1, \ldots, n$, we can define the pseudo-response $\widetilde{y}_i^{(m)}$ for subject $i$ at the $m$-{th} iteration as:
\begin{equation}\label{rapls:cox}
\widetilde{y}^{(m)}_i = \int_{\mathcal{S}} x_i(s)b^{(m)}_{1,p_1}(s)ds + \{\mu_i^{(m)}\}^{-1}(\Delta_i - \mu_i^{(m)}).
\end{equation}
We provide the details for deriving (\ref{rapls:cox}) in the supporting information. Next, we consider the following model:
\begin{equation}\label{rapls:cox:1}
\{\mu_i^{(m)}\}^{1/2}\widetilde{y}_i^{(m)} = \{\mu_i^{(m)}\}^{1/2}\int_{\mathcal{S}} x_i(s)b_1(s)ds + \delta_i^{(m)}, ~~ \rmn{for}~~ i = 1, \ldots, n.
\end{equation}
Note that (\ref{rapls:cox:1}) has the same form as (\ref{RAPLS7}), and thus we calculate the top $p_1$ RAPLS basis functions for (\ref{rapls:cox:1}), denoted by $\zeta^{(m)}_1(s), \ldots, \zeta^{(m)}_p(s)$. Similar to Step 1, model (\ref{intro:model:jm:surv}) can be approximated by
\begin{equation}\label{jm:surv:step2}
\lambda_i(t|\bmath  \omega_i, x_i(s), \bmath u_i)=  \lambda_0(t)\exp\left(\bmath  \omega_i^T\bmath{\gamma} + \sum_{j=1}^{p_1} b_{1,j} \left(\int_{\mathcal{S}} x_i(s) \zeta_j^{(m)}(s)ds \right) + \alpha m_i(t)\right), 
\end{equation}
and $b_1(s)$ can be approximated by $\sum_{j=1}^{p_1} b_{1,j} \zeta_j^{(m)}(s)$.

\textbf{Step 3:} Note that  FJM (\ref{intro:model:jm:long}) and (\ref{intro:model:jm:surv}) are now, respectively, approximated by the standard joint model with low-dimensional covariates (\ref{jm:long:step1}) and (\ref{jm:surv:step2}). Due to the unobserved random effects $\bmath u_i$, we propose to fit (\ref{jm:long:step1}) and (\ref{jm:surv:step2}) by the EM algorithm \citep{dempster1977maximum}. The details of the EM algorithm are given in the supporting information.
Let $\bmath\Theta_{p_0, p_1}^{(m + 1)}$, $\Lambda_{0, p_0, p_1}^{(m + 1)}(\cdot)$, $b_{0,j}^{(m+1)}$ and $b_{1,j}^{(m+1)}$ denote the EM estimates of $\bmath \Theta, \Lambda_0(\cdot), b_{0,j}$ and $b_{1,j}$, respectively. Then, $b^{(m+1)}_{0,p_0}(s)$ and $b^{(m+1)}_{1,p_1}(s)$ are, respectively, given by
\begin{equation}\label{fpls:update}
b^{(m+1)}_{0,p_0}(s) = \sum_{j=1}^{p_0} b_{0,j}^{(m+1)} \psi_j^{(m)}(s) ~~\mbox{and}~~ b^{(m+1)}_{1,p_1}(s) = \sum_{j=1}^{p_1} b_{1,j}^{(m+1)} \zeta_j^{(m)}(s).
\end{equation}


With appropriate initial values, the proposed FPLS algorithm for FJM is done by iterating between Steps 1, 2 and 3 until $\int_{\mathcal{S}} \left(b_{0,p_0}^{(m + 1)}(s) - b_{0,p_0}^{(m)}(s)\right)^2 ds + \int_{\mathcal{S}} \left(b_{1,p_1}^{(m + 1)}(s) - b_{1,p_1}^{(m)}(s)\right)^2 ds$ is less than a given threshold $\kappa_0$, which is set to be $10^{-6}$ in the following numerical studies.

\section{Implementation}
\label{ss:comp}

\subsection{Convergence}
Based on our experience in numerical studies, we make several comments for facilitating the convergence of the proposed FPLS algorithm. 
\begin{enumerate}
    \item[(i)]  
   The stability of the proposed FPLS algorithm may suffer from the non-orthogonality of the RAPLS basis functions. Since the key idea of the proposed FPLS algorithm is to find the low-dimensional space spanned by the RAPLS basis functions, we suggest orthogonalizing the RAPLS basis functions in Steps 1 and 2 before running the EM algorithm in Step 3.  
    
    \item[(ii)] As in most regression problems, it may be helpful to put all covariates on comparable scales. In particular, singular Hessian matrix may appear during the EM iterations (Step 3) if the two integrals $\int_{\mathcal{S}} x_i(s) \psi_j^{(m)}(s)ds$ and $\int_{\mathcal{S}} x_i(s) \zeta_j^{(m)}(s)ds$ are vastly different from the scalar predictors. This may happen 
    in practice if $x_i(s)$ is not appropriately scaled because these two integrals are approximated by summing over millions of grid points. 
    We elaborate on the scaling process as follows. As suggested in (i), we can orthogonalize $\psi_j^{(m)}(s)$ and $\zeta_j^{(m)}(s)$ such that $\int_{\mathcal{S}} \psi_{j_1}^{(m)}(s)\psi_{j_2}^{(m)}(s)ds = \int_{\mathcal{S}} \zeta_{j_1}^{(m)}(s)\zeta_{j_2}^{(m)}(s)ds = I(j_1 = j_2)$.
    Then, by using the Cauchy–Schwarz inequality, it is easy to show that 
    \[|\int_{\mathcal{S}} \overline{x}(s) \psi_j^{(m)}(s)ds| \leq \|\overline{x}(s)\|_2 ~\mbox{and}~ |\int_{\mathcal{S}} \overline{x}(s) \zeta_j^{(m)}(s)ds| \leq \|\overline{x}(s)\|_2,\]
    where $\|\overline{x}(s)\|_2 = \left(\int_{\mathcal{S}} \overline{x}^2(s) ds \right)^{1/2}$.
    This implies that a sensible way to scale $x_i(s)$ is to multiply $x_i(s)$ by a constant such that $\|\overline{x}_i(s)\|_2$ has comparable size to the scalar predictors.

    \item[(iii)]  As for most iterative algorithms, an appropriate choice of initial values is critical for the convergence of the algorithm. 
    We use the FPCA estimates as the initial values for the proposed FPLS algorithm. The key to finding the FPCA estimates is estimating eigenfunctions. Unlike \cite{LiLuo2017}, which advocates the \texttt{fpca.sc} function in the \texttt{refund} package \citep{refundr} or \texttt{fpca.mle} and \texttt{fpca.score} functions in the \texttt{fpca} package \citep{fpcarpack} in {\tt R}, we use the method based on the singular value decomposition (SVD) proposed by \cite{zipunnikov2011functional}. The reason is that the two \texttt{R} packages advocated by \cite{LiLuo2017} can be computationally intensive with ultrahigh-dimensional images.
    For example,  to perform FPCA for a data set with 200 subjects, each having a $300 \times 300$ image (the simulation study in Section 4), \texttt{fpca.sc} and \texttt{fpca.mle} require at least $60$ Gb of RAM memory and take hours to run, whereas the SVD method only requires less than 6 Gb of RAM memory and finishes in seconds. Once eigenfunctions are estimated, one can follow the method in \cite{LiLuo2017}
    to obtain the FPCA estimators.
    \item[(iv)]  We also suggest a step of stochastic approximation to accelerate the convergence. Specifically, instead of using (\ref{fpls:update}) to obtain the updates $b_{0,p_0}^{(m+1)}(s)$ and $b_{1,p_1}^{(m+1)}(s)$, we consider 
    $b^{(m+1)}_{0,p_0}(s) = (1 - a^{(m)})b^{(m)}_{0,p_0}(s) + a^{(m)}\sum_{j=1}^{p_0} b_{0,j}^{(m+1)} \psi_j^{(m)}(s)$ and $b^{(m+1)}_{1,p_1}(s) = (1 - a^{(m)})b^{(m)}_{1,p_1}(s) + a^{(m)}\sum_{j=1}^{p_1} b_{1,j}^{(m+1)} \zeta_j^{(m)}(s).$ Here, $a^{(m)}$ is called the step size at the $m$-{th} iteration. As demonstrated in \cite{WALK1978430}, \cite{YIN1990116} and the references therein, flexible step sizes can accelerate the convergence of the algorithm as well as stabilize the performance. In our numerical studies, we use $a^{(m)} = 1/m$.
\end{enumerate}
To apply the proposed FPLS algorithm to 2D or 3D images, one can vectorize them in any way since the proposed algorithm is invariant with respect to arbitrary vectorization of the images. 
In our real data analysis where each image has more than 500,000 voxels, each iteration of the proposed algorithm takes less than 1 minute to run, and the entire algorithm becomes stable in a few steps.


\subsection{Choice of $p_0$ and $p_1$}
So far we have discussed the algorithm with predetermined $p_0$ and $p_1$. As demonstrated in \cite{LiLuo2017}, changing $p_0$ and $p_1$ can have a large effect on the estimation accuracy of the parameters in FJM. Generally, smaller $p_0$ and $p_1$ may lead to a larger bias, whereas larger $p_0$ and $p_1$ may lead to a larger variance. To balance bias and variance, we use the Bayesian information criterion (BIC) to choose $p_0$ and $p_1$. Specifically, let $\widehat{\bmath\Theta}_{p_0, p_1}, \widehat{b}_{0, p_0}(s), \widehat{b}_{1,p_1}(s)$ and $\widehat{\Lambda}_{0,p_0,p_1}(\cdot)$ be the estimates of $\bmath \Theta, b_0(s), b_1(s)$ and $\Lambda_0(\cdot)$.
Then, the BIC statistic is defined as
\[\mbox{BIC}(p_0, p_1) = \mbox{log}(n)(p_0 + p_1) - 2\int \mbox{log}(L_{n,\mbox{com}}(\widehat{\bmath\Theta}_{p_0, p_1}, \widehat{b}_{0, p_0}(s), \widehat{b}_{1,p_1}(s),\widehat{\Lambda}_{0,p_0,p_1}(\cdot)))d\bmath u_1\cdots\bmath u_n,\]
where $L_{n,\mbox{com}}(\widehat{\bmath\Theta}_{p_0, p_1}, \widehat{b}_{0, p_0}(s), \widehat{b}_{1,p_1}(s), \widehat{\Lambda}_{0,p_0,p_1}(\cdot))$ is the complete data likelihood, of which the explicit from is given in the supporting information.
Numerically, we can use a grid search to find the optimal $p_0$ and $p_1$ that minimize BIC$(p_0, p_1)$.

\section{Simulation Study}
\label{s:simu}
\begin{figure}
\centerline{\includegraphics[width= \textwidth]{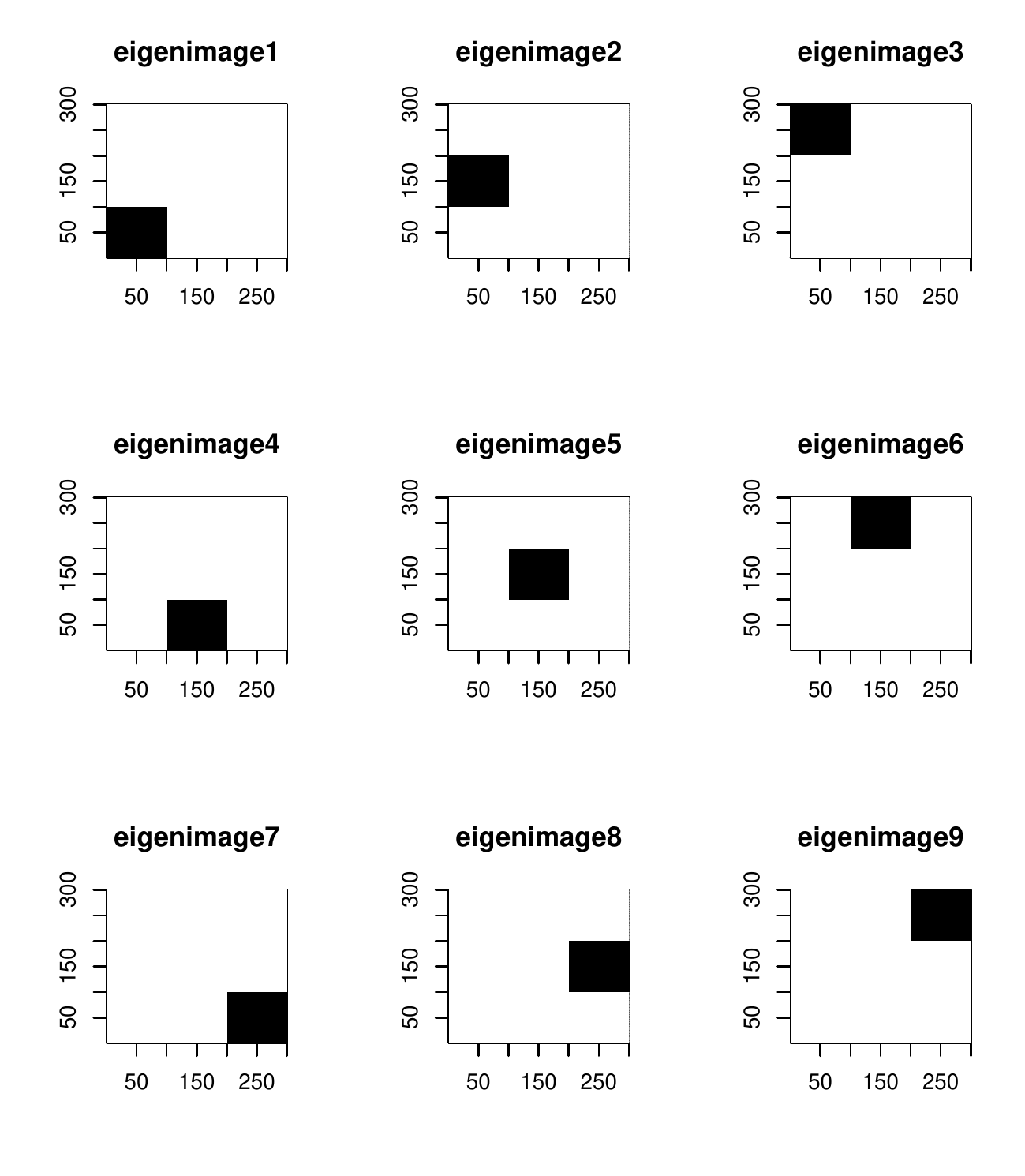}}
\caption{True greyscale eigenimages used in the simulation study.}
\label{eigenimage}
\end{figure}

In this section, we carry out a simulation study to compare the proposed FPLS algorithm with  FPCA for FJM (\ref{intro:model:jm:long}) and (\ref{intro:model:jm:surv}). 
For $ i = 1, \ldots, n$, we first independently simulated $X_i(s)$ according to  the generating process  
\[X_i(s)= \sum_{k=1}^{9}k^{-1/4}\xi_{ik}\phi_k(s), \xi_{ik} \stackrel{i.i.d}{\sim} N(0,1), s \in \mathcal{S},\] 
where eigenimages $\phi_k(s)$ are displayed in Fig. \ref{eigenimage} and $\mathcal{S} = [1,300] \times [1,300]$. These eigenimages can be thought of as 2D greyscale images with pixel intensities on the $[0,1]$ scale. The black pixels are set to 1 and the white ones are set to 0. 
We generated $Z_i$ from a normal distribution with mean 0 and variance $\xi_{i2}^2/9$. This indicates that  $Z_i$ and $X_i(s)$ are correlated.
We denote $m_i(t) = \beta_0 + \beta_1t_{ij}+Z_i\beta_2+ \int_{\mathcal{S}} X_i(s)b_0(s)ds + u_i$, where $u_i \stackrel{i.i.d}{\sim} N(0,1)$. 
Then, the true event time $T_i$ was generated based on the proportional hazards model as follows:
\begin{equation}\label{JM:simu:surv}
\lambda_i(t|Z_i,X_i(s),u_i) = \rmn{exp}\{\alpha m_i(t) + \int_{\mathcal{S}} X_i(s)b_1(s)ds + Z_i\gamma\}.
\end{equation}
Next, we independently generated the censoring time $C_i$ from a uniform distribution $ U(0, c_0) $, where  $ c_0 > 0 $ controls the censoring rate. Instead of observing both $T_i$ and $C_i$, we only observed $\widetilde{T}_i = \rmn{min}(T_i, C_i)$ and $\Delta_i = I(T_i \leq C_i)$.  Next, the longitudinal follow up time $\{t_{ij}\}_{j=1,2,3}$ was simulated from a uniform distribution in $(0, \widetilde{T}_i)$, as only the measurements that are collected before the observed survival time can be used to predict the survival event. Then, the longitudinal outcome $y_{ij}$ was generated by
\[y_i(t_{ij}) = m_i(t_{ij}) + \epsilon_{ij},\]
where $\epsilon_{ij} \stackrel{i.i.d}{\sim} N(0,0.4^2)$.
We set $\beta_0 = 0.7, \beta_1 = 1, \beta_2 = 2, \alpha = 2$ and $\gamma = 2$, and consider two scenarios of $b_0(s)$ and $b_1(s)$: (i) $b_0(s) =  b_1(s) = \sum_{k=1}^{5}k^{-3/2}\phi_k(s)$ and (ii) $b_0(s) = \sum_{k=5}^{9}(k-4)^{-1/2}\phi_k(s), b_1(s) = \sum_{k=1}^{5}k^{-1/2}\phi_k(s)$. 

We consider $n = 200$ and $500$. The $c_0$ is selected for each scenario to yield a censoring rate of 60\% that mimics our real data analysis. 
For each pair of $n$ and $c_0$, we generated $1000$ independent data sets. For simplicity, here we consider the same number of basis functions to estimate $b_0(s)$ and $b_1(s)$, i.e., $p_0 = p_1 = p$. In practice, $p_0$ and $p_1$ can be different; this will be considered in the real data analysis in Section 5. The optimal $p$ was then selected by the BIC statistic, given in Section 3.2.
For the implementation of both FPLS and FPCA, generated images $X_i(s)$'s were unfolded to obtain vectors of size $d = 300 \times 300 = 90000$. As mentioned in Section 3.1, our implementation of FPCA is different from that in \cite{LiLuo2017}, and the FPLS algorithm uses the FPCA estimates as the initial values. 

We examined the mean squared error (MSE) for both functional parameters according to $\rmn{MSE}_{b_j} = \int_{\mathcal{S}} (\widehat{b}_j(s) - b_j(s))^2 ds$, where $\widehat{b}_j(s)$ is the FPCA or FPLS estimator of $b_j(s)$ for $j = 0$ and 1.
Fig. 3 shows results for scenario (i). In this scenario, both $b_0(s)$ and $b_1(s)$ are only informed by the top 5 eigenimages. The weight of the $j$-th eigenimage,
$j^{-1.5}$, decreases fast as $j$ increases; this further favors FPCA. It can be seen that the proposed FPLS algorithm performs comparably to FPCA: FPLS yields a more accurate estimate of $b_0(s)$, whereas FPCA performs slightly better in terms of estimating $b_1(s)$. This is sensible because the IRLS procedure used for the survival model may introduce more error. In scenario (ii) that does not favor FPCA, as shown in Fig. 4, the proposed FPLS algorithm has considerably higher estimation accuracy than FPCA for both $b_0(s)$ and $b_1(s)$. In particular, It can be seen from Fig. 4C that our FPLS algorithm yields accurate estimation of $b_0(s)$, whereas FPCA yields an MSE over 2. A straightforward calculation yields that $\int_{\mathcal{S}} b^2_0(s) ds = 2.28$, indicating that FPCA can barely capture any information from $b_0(s)$.  To see the reason clearly, we first note that $b_0(s)$ lies
in the span of the fifth to the ninth eigenimages. Unfortunately, these eigenimages cannot be consistently estimated in such an ultrahigh-dimensional setting ($d \gg n$). Thus, BIC tends to select small $p$ ($p < 5$) for FPCA, indicating that the resulting FPCA estimator is very close to 0. In contrast, since the FPLS basis functions incorporate information from the outcome, which contains information on $b_0(s)$, the FPLS estimator can still yield an accurate estimate of $b_0(s)$ regardless of the inconsistent estimates of the eigenimages. In practice, since we never know how the functional predictor informs the outcome, the proposed FPLS algorithm may be a more robust and accurate prediction tool, especially in neuroimaging studies with high-dimensional images and limited sample size.


\begin{figure}
\centerline{\includegraphics[width= \textwidth]{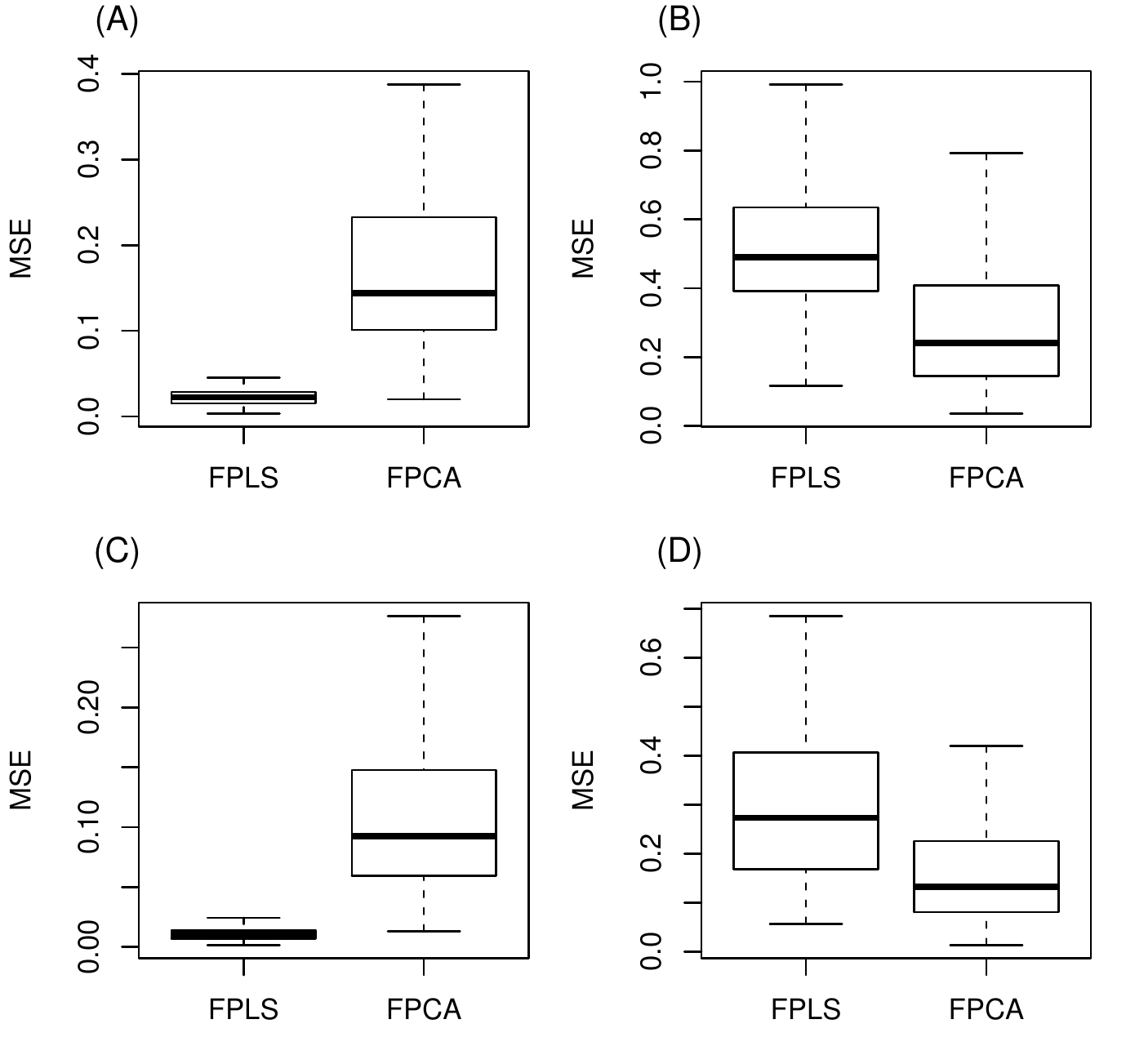}}
\caption{The boxplot of the $\mbox{MSE}$ for scenario (i) over 1000 replications: (A) $n = 200$ and $\mbox{MSE}_{b_0}$; (B): $n = 200$ and $\mbox{MSE}_{b_1}$; (C) $n = 500$ and $\mbox{MSE}_{b_0}$; (D): $n = 500$ and $\mbox{MSE}_{b_1}$.}
\label{scenario:1}
\end{figure}

\begin{figure}
\centerline{\includegraphics[width= \textwidth]{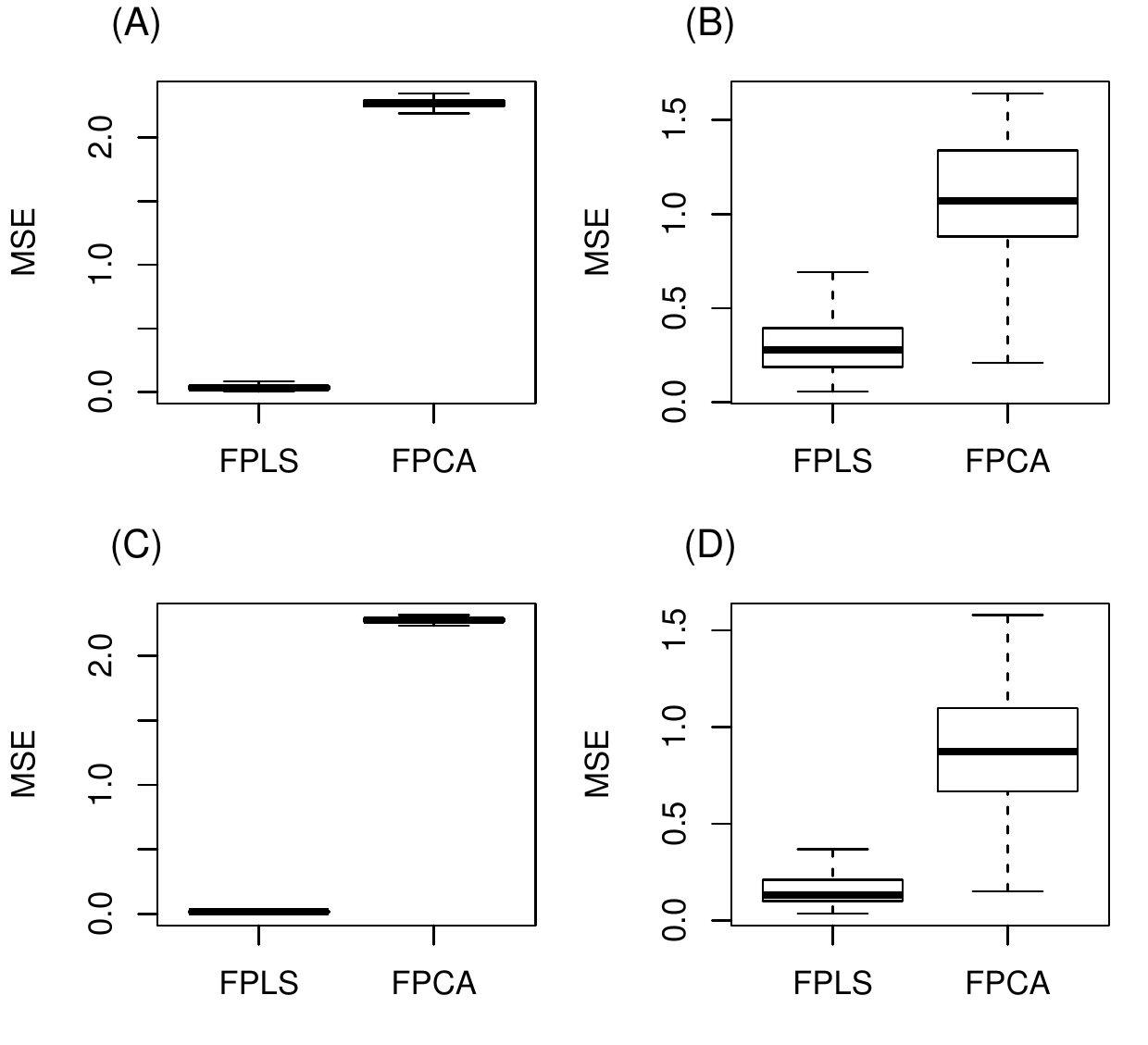}}
\caption{The boxplot of the $\mbox{MSE}$ for scenario (ii) over 1000 replications: (A) $n = 200$ and $\mbox{MSE}_{b_0}$; (B): $n = 200$ and $\mbox{MSE}_{b_1}$; (C) $n = 500$ and $\mbox{MSE}_{b_0}$; (D): $n = 500$ and $\mbox{MSE}_{b_1}$.}
\label{scenario:2}
\end{figure}

\section{Analysis of ADNI Data}

In this section, we jointly model the longitudinal trajectory of ADAS-Cog score and the time of conversion from MCI to AD using the selected 236 subjects. Specifically, we consider
\[y_{ij} =  m_i(t_{ij}) + \epsilon_{ij} ~~ \rmn{with}\]
\[m_i(t_{ij}) =  \beta_0 + {\bf z}_i^T \bmath \beta_1 + \beta_2 t_{ij} + \int_{\mathcal{S}}x_i(s)b_0(s)ds + u_i\] 
and 
\[\lambda_i(t|{\bf z}_i,x_i(s)) = \mbox~  \lambda_0(t)\exp\left({\bf z}_i^T \bmath \gamma + \int_{\mathcal{S}}x_i(s)b_1(s)ds + \alpha m_i(t)\right).\]
Here, $t_{ij}$ is the follow-up time for the $i$-{th} subject at the $j$-{th} visit and $y_{ij}$ is the ADAS-Cog score of the $i$-{th} subject at the $j$-{th} visit. The scalar covariate vector ${\bf z}_i$ includes  gender (1=male; 0=female), handedness (1=right; 0=left), and age at the first MCI diagnosis. The functional predictor $x_i(s)$ is the PET imaging data measured on $160\times 160 \times 96$ voxels. PET  directly measures the regional use of glucose, which indirectly reflects the brain activity of different brain regions.  The PET images we used here underwent four preprocessing steps, which are introduced in detail in the supporting information. We also removed the background regions outside the skull, so around 500,000 voxels remained. 
The parameter $\alpha$ links the two models. If $\alpha$ is nonzero, then there may be an unobserved association between the longitudinal and survival outcome. The random intercept $u_i$ is assumed to be normally distributed with mean 0 and variance $\sigma_u^2$. Given $u_i$, $\epsilon_{ij}$ follows a normal distribution with mean 0 and variance $\sigma^2_\epsilon$. The implementation of FPCA and FPLS is the same as that in Section 4.

\begin{figure}
\centerline{\includegraphics[width= \textwidth]{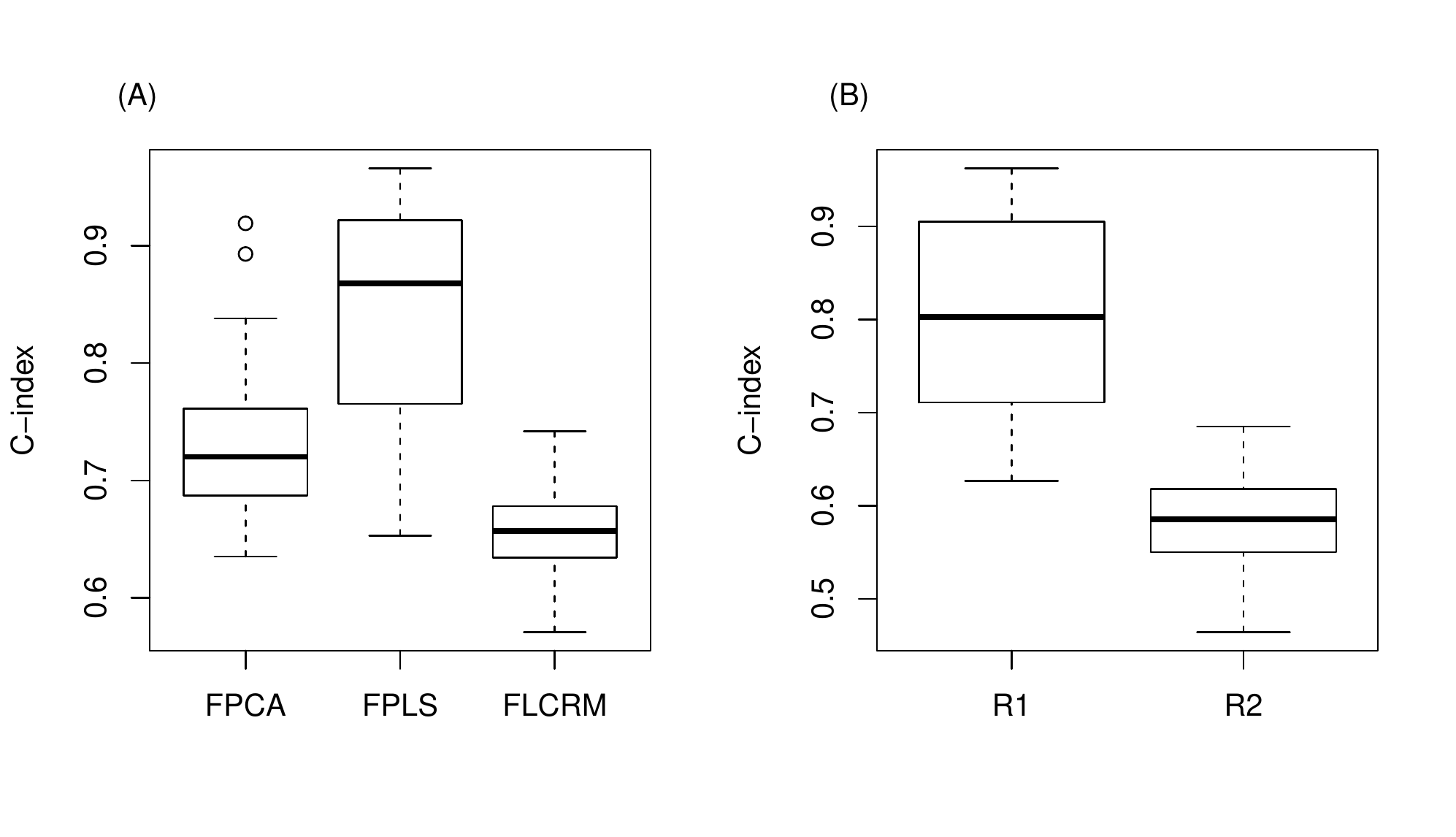}}
\caption{Boxplots of the C-index over 100 replications for the FPLS, FPCA, FLCRM, R1 and R2.}
\label{cindex:fig}
\end{figure}

We first used cross validation to compare the proposed FPLS algorithm with FPCA in terms of prediction accuracy of the survival time. To examine the predictive value of the longitudinal ADAS-Cog scores, we also predicted the survival time based on the following functional linear Cox regression model (FLCRM, \citealp{kong2018}): 
 \[\lambda_i(t|{\bf z}_i,x_i(s)) = \mbox~  \lambda_0(t)\exp\left({\bf z}_i^T \bmath \gamma + \int_{\mathcal{S}}x_i(s)b_1(s)ds\right).\]
 The prediction accuracy was examined according to the concordance index (C-index, \citealp{Harrell1996}), which can be calculated using the function  \texttt{"concordance.index()"} in the {\tt R} package \texttt{"survcomp"} \citep{survcomp2011}.  More specifically, we randomly selected 118 subjects as the training set and the remaining 118 subjects form the test set. We repeated this procedure 100 times. For each method in each replication, we fitted the FJM using the training set with the optimal number of basis functions, i.e. $p_0$ and $p_1$, selected by BIC, and then computed the C-index using the test set. Inspecting Fig. \ref{cindex:fig}A shows that FPLS yields higher prediction accuracy than FPCA. Moreover, both FPLS and FPCA substantially outperform FLCRM, demonstrating that the ADAS-Cog score may be an important predictor of the progression from MCI to AD.

To examine the predictive value of  PET imaging data, two reduced models are considered. The first reduced model excludes the imaging predictor from the longitudinal model, and the second one excludes the imaging predictor from the survival model, denoted by R1 and R2, respectively. Using the same training and testing data sets, we fitted these two reduced models using our FPLS algorithm with the optimal $p_0$ or $p_1$ selected by BIC and calculated the C-index. Fig.  \ref{cindex:fig}B shows that two reduced models yield lower prediction accuracy than the FPLS in Fig. \ref{cindex:fig}A, demonstrating the predictive value of the PET imaging data in terms of jointly predicting ADAS-Cog scores and the progression to AD. Moreover, as we aim at predicting the survival outcome, the prediction accuracy suffers more when the imaging predictor is removed from the survival model. 

\begin{figure}
\centerline{\includegraphics[width = 1.2\textwidth,height = 0.5 \textheight]{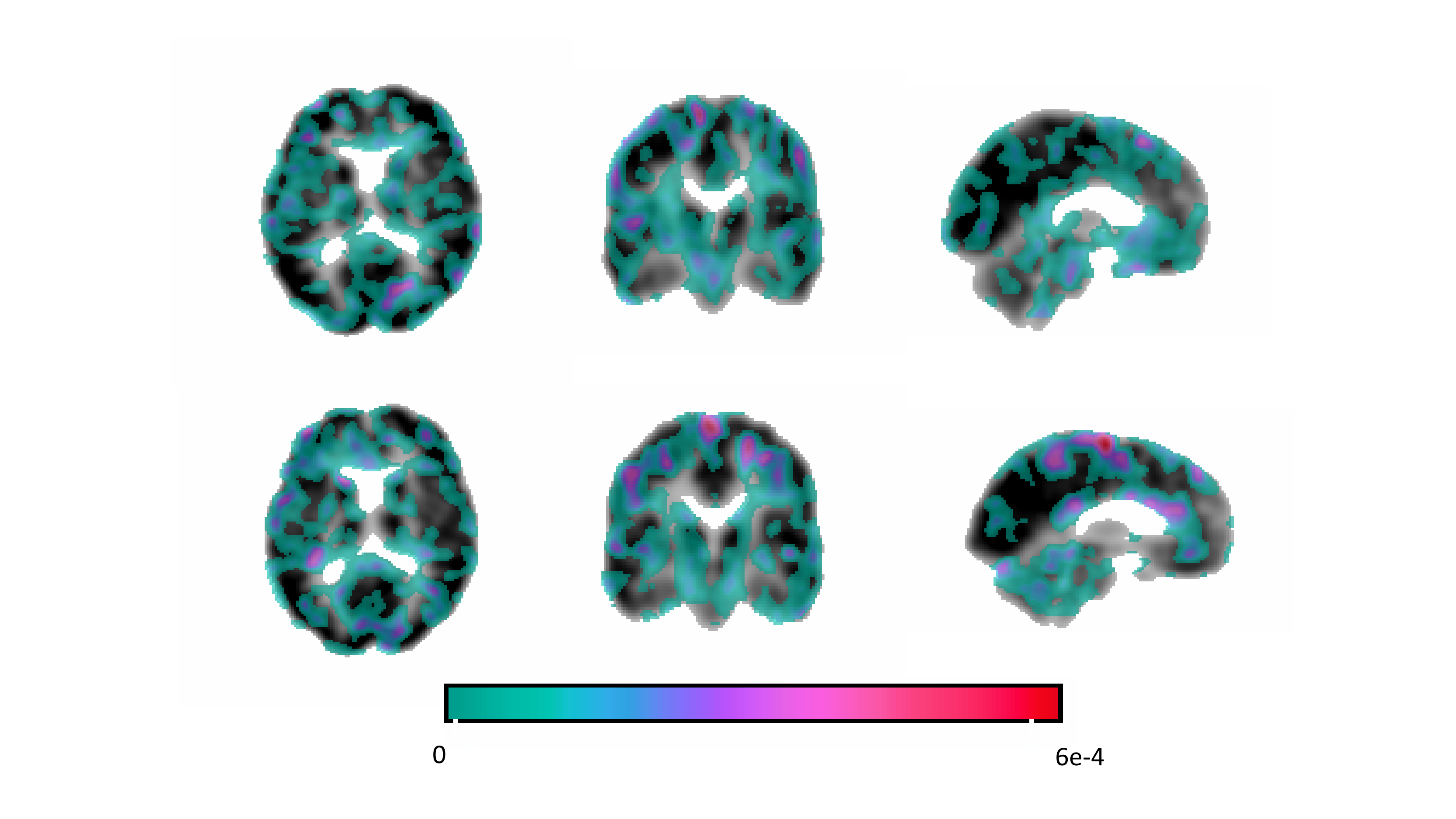}}
\caption{Positive regions of the estimated coefficient images obtained from the FPLS algorithm with $p_0 = 18$ and $p_1 = 10$: The top and bottom rows display the positive regions of the estimated $b_0(s)$ and $b_1(s)$ respectively. From left to right, each coefficient is displayed in the views of transverse, coronal and sagittal planes. The slices are located at (48,80,80).}\label{fbeta:JM}
\end{figure}

Finally, we considered the complete cohort of 236 subjects to estimate the unknown parameters using our FPLS algorithm. 
The optimal $p_0$ and $p_1$ selected by BIC are 18 and 10,  respectively. 
The estimated $\alpha$ is 2.8, indicating that patients with higher ADAS-Cog score may be more likely to progress to AD.
Fig. \ref{fbeta:JM} displays the positive regions of the estimates of $b_0(s)$ and $b_1(s)$. It can be seen that several functional regions over the brain, such as the hippocampus, frontal lobe and temporal horn of the lateral ventricle are identified to be positively associated with the progression from MCI to AD.



\section{Discussions}

In this article, we developed a novel FPLS algorithm for the estimation and prediction of FJM. 
We examined its performance in simulation studies and an application to ADNI. As shown in the numerical studies, FPLS can yield comparable results to FPCA in the setting that strongly favors FPCA, whereas it yields accurate estimates in the setting where FPCA completely fails. 
Hence, the proposed FPLS algorithm may be a more robust and powerful prediction tool for FJM than FPCA when massive neuroimaging data are involved. 

It should be noted, however, that it is very challenging to theoretically derive (asymptotic) confidence intervals of any FPLS based estimators. The reason may be inherent to the key feature of FPLS; that is, the FPLS basis functions depend on the outcome. Consequently, the design matrix in FPLS regressions also involves the error term.
Thus, bootstrapping methods have been suggested for constructing confidence intervals of PLS type of estimators \citep{wold2001pls}. Further studies of the inferential problems associated with the proposed FPLS algorithm may be a fruitful area of future research.

\backmatter

\section*{Acknowledgement}

Dr. Zhu's work was partially supported by NIH grants R01MH086633  and R01MH116527. The content is solely the responsibility of the authors and does not necessarily represent the official views of the NIH. Data used in the preparation of this article were obtained from the Alzheimers Disease Neuroimaging Initiative (ADNI) database (adni.loni.usc.edu). As such, the investigators within the ADNI contributed to the design and implementation of ADNI and/or provided data but did not participate in analysis or writing of this report. A complete listing of ADNI investigators can be found at http://adni.loni.usc.edu/wpcontent/uploads/how to apply/ADNI Acknowledgement List.pdf.


%
 \bibliographystyle{biom} 
\bibliography{biomtemplate.bib}

\section*{Supporting Information}

Web Appendices, Tables and Figures referenced in Sections 2, 3, and 4 are available with
this paper at the Biometrics website on Wiley Online
Library. Our code and example data for implementing the FPLS algorithm are also available at the Biometrics website on Wiley Online
Library.
\vspace*{-8pt}








\label{lastpage}

\end{document}